\documentclass[letterpaper,12pt]{article}
\pdfoutput=1

\usepackage{amsmath}
\usepackage{diagbox}
\usepackage{jheppub}
\usepackage{multirow}

\DeclareMathOperator{\Gr}{Gr}
\DeclareMathOperator{\sign}{sign}
\newcommand{\an}[1]{\langle#1\rangle}

\allowdisplaybreaks[1]

\title{\large Yangian Invariants and Cluster Adjacency in $\mathcal{N}=4$ Yang-Mills}

\author{Jorge Mago,$^1$}
\author{Anders Schreiber,$^1$}
\author{Marcus Spradlin$^{2}$}
\author{and Anastasia Volovich$^1$}

\affiliation{$^1$ Department of Physics, Brown University, Providence, RI 02912, USA}

\affiliation{$^2$ Department of Physics and Brown Theoretical Physics Center, Brown University, Providence, RI 02912, USA}

\abstract{We conjecture that every rational Yangian invariant in $\mathcal{N}=4$ SYM theory satisfies a recently introduced notion of cluster adjacency. We provide evidence for this conjecture by using the Sklyanin Poisson bracket on $\Gr(4,n)$ to check numerous examples.}

\begin{document}

\maketitle

\section{Introduction}

In recent years cluster algebras have shed interesting light on the mathematical properties of scattering amplitudes in planar $\mathcal{N}=4$ supersymmetric Yang-Mills (SYM) theory~\cite{Golden:2013xva}. Cluster algebraic structure manifests itself in several distinct ways, notably including the appearance of certain $\Gr(4,n)$ cluster coordinates in the symbol alphabets~\cite{Golden:2013xva, Golden:2013lha, Golden:2014pua, DelDuca:2016lad}, cobrackets~\cite{Golden:2013xva, Golden:2014xqa, Golden:2014xqf, Harrington:2015bdt, Golden:2018gtk}, and integrands~\cite{ArkaniHamed:2012nw} of $n$-particle amplitudes.

There has been a recent revival of interest in the cluster structure of SYM amplitudes following the observation~\cite{Drummond:2017ssj} that certain amplitudes exhibit a property called \emph{cluster adjacency}. Cluster coordinates are grouped into sets called clusters, with two coordinates being called adjacent if there exists a cluster containing both. The central problem of the ``cluster adjacency'' literature is to identify (and, hopefully, to explain!) correlations between sets of pairs (or larger groupings) of cluster coordinates, and the manner in which those pairs are observed to appear together in various amplitudes.

For example, for loop amplitudes, all evidence available to date~\cite{Drummond:2018dfd, CaronHuot:2011ky, CaronHuot:2011kk,Dixon:2014iba, Drummond:2014ffa, Dixon:2015iva, Caron-Huot:2016owq, Dixon:2016nkn, Dixon:2016apl, Drummond:2018caf, Golden:2019kks, Caron-Huot:2019vjl} supports the hypothesis that two cluster coordinates appear in adjacent symbol entries only if they are cluster adjacent. In~\cite{Golden:2019kks} it was shown that this type of cluster adjacency implies the Steinmann relations~\cite{Steinmann,Steinmann2,Cahill:1973qp}. For tree amplitudes a somewhat analogous version of cluster adjacency was proposed in~\cite{Drummond:2018dfd}, where it was checked in several cases, and conjectured in general, that every Yangian invariant in the BCFW expansion of tree-level amplitudes in SYM theory has poles given by cluster coordinates that are all contained in a common cluster.

In this paper we provide further evidence for this and the even stronger conjecture that cluster adjacency holds for every rational Yangian invariant in SYM theory, even those that do not appear in any representation of tree amplitudes.

In Sec.~2 we review the main tool of our analysis, the Sklyanin Poisson bracket~\cite{Sklyanin:1982tf,GSV} which can be used to diagnose whether two cluster coordinates on $\Gr(4,n)$ are adjacent, which we will call the \emph{bracket test}~\cite{Golden:2019kks}. In Sec.~3 we review the Yangian invariants of SYM theory and explain how (in principle) to use the bracket test to provide evidence that N${}^k$MHV Yangian invariants satisfy cluster adjacency. We carry out this check for all $k \le 2$ invariants and many $k=3$ invariants.

Before proceeding we make a few comments clarifying the ways in which our tests are weaker than the analysis of~\cite{Drummond:2018dfd}, and the ways in which they are stronger:
\begin{enumerate}
\setlength\itemsep{0em}
\item{It could have happened that only certain representations of tree-level amplitudes (depending, perhaps, on the choice of shifts during intermediate steps of BCFW recursion) satisfy cluster adjacency, but as already noted, our results suggest that every rational Yangian invariant satisfies cluster adjacency. If true, this suggests that the connection between cluster adjacency and Yangian invariants admits a mathematical explanation independent of the physics of scattering amplitudes.}
\item{For any fixed $k$ there are finitely many functionally independent N${}^k$MHV Yangian invariants. If it is known that these all satisfy cluster adjacency, it immediately follows that the $n$-particle N${}^k$MHV amplitude satisfies cluster adjacency for all $n$. Our results therefore extend the analysis of~\cite{Drummond:2018dfd} in both $k$ and $n$.}
\item{However, unlike in~\cite{Drummond:2018dfd}, we make no attempt to check whether each of the polynomial factors we encounter is actually a $\Gr(4,n)$ cluster coordinate. Indeed for $n > 7$ there is no known algorithm for determining, in finite time, whether or not a given homogeneous polynomial in Pl\"ucker coordinates is a cluster coordinate. The bracket does not help here; it is trivial to write down pairs of polynomials that pass the bracket test but are not cluster coordinates.}
\item{In the examples checked in~\cite{Drummond:2018dfd} it was noted that each term in a BCFW expansion of an amplitude had the property that there exists a cluster of $\Gr(4,n)$ that simultaneously contains all of the cluster coordinates appearing in the denominator of that term. Our test is much weaker in that it can only establish pairwise cluster adjacency. For example, if we encounter a term with three polynomial factors $p_1$, $p_2$ and $p_3$, our test provides evidence that there is some cluster containing $p_1$ and $p_2$, and also some cluster containing $p_2$ and $p_3$, and also some cluster containing $p_1$ and $p_3$, but the bracket test cannot provide any evidence for or against the existence of a cluster simultaneously containing all three.}
\end{enumerate}

In this paper we rely on theoretical experimentation to find evidence in support of the cluster adjacency of Yangian invariants, but it is natural to hope that some kind of connection might be discovered that could shine light on a path towards a general proof.  This seems to be a difficult problem on two fronts: SYM theory admits a whole zoo of Yangian invariants (see Chapter~12 of~\cite{ArkaniHamed:2012nw} for a discussion of their classification), and the structure of $\Gr(4,n)$ is still rather mysterious for $n>7$ (when the algebra is infinite).  In particular, as noted above, there is no known algorithm for determining whether a given quantity is a cluster coordinate, let alone whether two coordinates are cluster adjacent.  Encouragingly, however, some progress towards a general proof has recently been made in~\cite{Lukowski:2019sxw}, which studied Yangian invariants in the so-called $m=2$ toy model of SYM theory.  There it was found to be possible to write down an explicit formula for arbitrary Yangian invariants, in a form that manifestly exhibits cluster adjacency with respect to the $\Gr(2,n)$ cluster algebra, whose simple structure is completely understood.

\section{Cluster Coordinates and the Sklyanin Poisson Bracket}

The objects of study in this paper will be certain rational functions on the kinematic space of $n$ cyclically ordered massless particles, of the type that appear in tree-level gluon scattering amplitudes. A point in this kinematic space is conveniently parameterized by a collection of $n$ \emph{momentum twistors}~\cite{Hodges:2009hk} $Z_1^I, \ldots, Z_n^I$, each of which can be regarded as a four-component $(I \in \{1,\ldots,4\}$) homogeneous coordinate on $\mathbb{P}^3$.

In these variables dual conformal symmetry~\cite{Drummond:2008vq} is realized by $SL(4, \mathbb{C})$ transformations. For a given collection of $n$ momentum twistors, the $\binom{n}{4}$ \emph{Pl\"ucker coordinates} are the $SL(4, \mathbb{C})$-invariant quantities
\begin{align}
\langle i\,j\,k\,l \rangle \equiv \epsilon_{IJKL} Z_i^I
Z_j^J Z_k^K Z_l^L\,.
\end{align}

The $\Gr(4,n)$ Grassmannian cluster algebra, whose structure has been found to underlie at least certain amplitudes in SYM theory, is a commutative algebra with generators called \emph{cluster coordinates}. Every cluster coordinate is a polynomial in Pl\"uckers that is homogeneous under a projective rescaling of each momentum twistor separately, for example
\begin{align}
\langle 1\,2\,6\,7 \rangle \langle 2\,3\,4\,5 \rangle
- \langle 1\,2\,4\,5 \rangle \langle 2\,3\,6\,7 \rangle\,.
\end{align}
Every Pl\"ucker coordinate is, on its own, a cluster coordinate. For $n < 8$ the number of cluster coordinates is finite and they can easily be enumerated, but for $n > 7$ the number of cluster coordinates is infinite.

The cluster coordinates of $\Gr(4,n)$ are grouped into non-disjoint sets of cardinality $4n{-}15$ called \emph{clusters}. Two cluster coordinates are said to be \emph{cluster adjacent} if there exists a cluster containing both. The $n$ Pl\"ucker coordinates $\langle 1\,2\,3\,4\rangle$, $\langle 2\,3\,4\,5 \rangle$, $\cdots$, $\langle n\,1\,2\,3\rangle$ containing four cyclically adjacent momentum twistors play a special role; these are called \emph{frozen} coordinates and are elements of every cluster. Therefore, each frozen coordinate is adjacent to every cluster coordinate.

Two Pl\"ucker coordinates are cluster adjacent if and only if they satisfy the so-called \emph{weak separation criterion}~\cite{LZ,OPS}. In order to address the central problem posed in the Introduction, it is desirable to have an efficient algorithm for testing whether two more general cluster coordinates are cluster adjacent. As proposed in~\cite{Golden:2019kks}, the Sklyanin Poisson bracket~\cite{Sklyanin:1982tf,GSV} $\{\, , \}$ can serve because of the expectation (not yet completely proven, as far as we are aware) that two cluster coordinates $a_1$, $a_2$ are adjacent if and only if $\{ \log a_1, \log a_2 \} \in \frac{1}{2} \mathbb{Z}$.

In the next section we use the Sklyanin Poisson bracket to test the cluster adjacency properties of Yangian invariants. To that end let us briefly review, following~\cite{Golden:2019kks} (see also~\cite{Vergu:2015svm}) how it can be computed. First, any generic $4 \times n$ momentum twistor matrix $Z_i^I$ can be brought into the gauge-fixed form
\begin{align}
\label{eqn:zmatrix2}
Z_i^I=
\begin{pmatrix}
1 & 0 & 0 & 0 & y^1{}_5 & \cdots & y^1{}_n \\
0 & 1 & 0 & 0 & y^2{}_5 & \cdots & y^2{}_n \\
0 & 0 & 1 & 0 & y^3{}_5 & \cdots & y^3{}_n \\
0 & 0 & 0 & 1 & y^4{}_5 & \cdots & y^4{}_n \\
\end{pmatrix}
\end{align}
by a suitable $GL(4, \mathbb{C})$ transformation. The Sklyanin Poisson bracket of the $y$'s is defined as
\begin{align}
\label{eqn:sklybrack1}
\{ y^I{}_a, y^J{}_b \} &= \frac{1}{2} (\text{sign} (J-I) - \text{sign} (b-a)) y^J{}_a y^I{}_b\,.
\end{align}
Finally, the Sklyanin Poisson bracket of two arbitrary functions $f$, $g$ of momentum twistors can be computed by plugging in the parameterization~\eqref{eqn:zmatrix2} and then using the chain rule
\begin{align}
\label{eqn:sklychainrule}
\{ f(y) , g(y) \} &= \sum_{a,b =1}^n \sum_{I, J = 1}^4 \frac{\partial f}{\partial y^I{}_a} \frac{\partial g}{\partial y^J{}_b} \{ y^I{}_a, y^J{}_b \}\,.
\end{align}

\section{An Adjacency Test for Yangian Invariants}

The conformal~\cite{Sohnius:1981sn} and dual conformal symmetry of scattering amplitudes in SYM theory combine to generate a Yangian~\cite{Drummond:2009fd} symmetry. \emph{Yangian invariants}~\cite{Drummond:2008vq, Mason:2009qx, ArkaniHamed:2009vw, ArkaniHamed:2009dg, ArkaniHamed:2009sx, Drummond:2010uq, Ashok:2010ie, ArkaniHamed:2012nw, Drummond:2010qh} are the basic building blocks in terms of which amplitudes can be constructed. We say that a Yangian invariant\footnote{Importantly, throughout this paper it should be understood that whenever we say ``Yangian invariant'' we mean ``\emph{positive} Yangian invariant,'' in the spirit of~\cite{ArkaniHamed:2012nw}.} is \emph{rational} if it is a rational function of momentum twistors; equivalently, it has intersection number $\Gamma = 1$ in the terminology of~\cite{ArkaniHamed:2012nw,Bourjaily:2012gy}. Any $n$-particle tree-level amplitude in SYM theory can be written as the $n$-particle Parke-Taylor-Nair superamplitude~\cite{Parke:1986gb,Nair:1988bq} times a linear combination of rational Yangian invariants (see for example~\cite{Drummond:2008cr}). In general the linear combination is not unique since Yangian invariants satisfy numerous linear relations.

Yangian invariants are actually superfunctions: an $n$-particle invariant is a polynomial of uniform degree $4k$ in $4kn$ Grassmann variables $\chi_i^A$, where $k$ is the N${}^k$MHV degree. For a rational Yangian invariant $Y$, the coefficient of each distinct term in its expansion in $\chi$'s can be uniquely factored into a ratio of products of polynomials in Pl\"ucker coordinates, with each polynomial having uniform weight in each momentum twistor separately. Let $\{p_i\}$ denote the union of all such polynomials that appear in the denominator of the expansion of $Y$. Then we say that $Y$ \emph{passes the bracket test} if
\begin{align}
\label{eqn:testwedo}
\Omega_{ij} \equiv \{ \log p_i, \log p_j \} \in \frac{1}{2} \mathbb{Z} \quad \forall i,j\,.
\end{align}

As explained in~\cite{ArkaniHamed:2012nw}, $n$-particle Yangian invariants can be classified in terms of permutations on $n$ elements. Since the bracket test is invariant\footnote{\label{ftn:fncyclic}Certainly the \emph{value} of the Sklyanin Poisson bracket is not in general cyclic invariant, since evaluating it requires making a gauge choice which breaks cyclic symmetry, such as in~\eqref{eqn:zmatrix2}, but the binary statement of whether some pair does or does not have half-integer valued bracket is cyclic invariant.} under the $\mathbb{Z}_n$ cyclic group that shifts the momentum twistors $Z_i \to Z_{i+1\,{\rm mod}\,n}$, we only need to consider one member from each cyclic equivalence class. The number of cyclic classes of rational N${}^k$MHV Yangian invariants with nontrivial dependence on $n$ momentum twistors was tabulated for various $k$ and $n$ in Table~3 of~\cite{ArkaniHamed:2012nw}. We record these numbers here, correcting typos in the $(3,15)$ and $(4,20)$ entries:
\vspace{-0.5cm}
\begin{center}
{\tiny
\begin{tabular}{c|cccccccccccccccc|c}
\diagbox{$k$}{$n$}
& 5 & 6 & 7 & 8 & 9 & 10 & 11 & 12 & 13 & 14 & 15 & 16 & 17 & 18 & 19 & 20 & {\bf Total} \\
\hline
1 & 1 & 0 & 0 & 0 & 0 & 0 & 0 & 0 & 0 & 0 & 0 & 0 & 0 & 0 & 0 & 0 & {\bf 1} \\
2 & 0 & 1 & 2 & 5 & 4 & 1 & 0 & 0 & 0 & 0 & 0 & 0 & 0 & 0 & 0 & 0 & {\bf 13} \\
3 & 0 & 0 & 1 & 6 & 54 & 177 & 298 & 274 & 134 & 30 & 3 & 0 & 0 & 0 & 0 & 0 & {\bf 977} \\
4 & 0 & 0 & 0 & 1 & 13 & 263 & 1988 & 7862 & 18532 & 28204 & 28377 & 18925 & 8034 & 2047 & 270 & 17 & {\bf 114533}
\end{tabular}
}
\end{center}

When they appear in scattering amplitudes, Yangian invariants typically have trivial dependence on several of the particles. For example the five-particle NMHV Yangian invariant $Y^{(1)}(Z_1, Z_2, Z_3, Z_4, Z_5)$ could appear in a nine-particle NMHV amplitude as $Y^{(1)}(Z_2, Z_4, Z_5, Z_7, Z_8)$, among other possibilities. Fortunately, because of the simple $\sign(b-a)$ dependence on column number in the definition~\eqref{eqn:sklybrack1}, the bracket test is insensitive to trivial dependence on additional momentum twistors\footnote{As in footnote~\ref{ftn:fncyclic}, the actual value of the Sklyanin Poisson bracket will in general change if the particle relabeling affects any of the first four gauge-fixed columns of $Z$.}.

Therefore for any fixed $k$, but arbitrary $n$, we can provide evidence for the cluster adjacency of every rational $n$-particle N${}^k$MHV Yangian invariant by applying the bracket test described above~\eqref{eqn:testwedo} to each one of the (finitely many) rational Yangian invariants. In the next few subsections we present the results of our analysis, beginning with the trivial but illustrative case of $k=1$.

\subsection{NMHV}

The unique $k=1$ Yangian invariant is the well-known \emph{five-bracket}~\cite{Mason:2009qx} (originally presented as an ``$R$-invariant'' in~\cite{Drummond:2008vq})
\begin{align}
\label{eqn:nmhvinvariant}
Y^{(1)} = [1,2,3,4,5] \equiv \frac{\delta^{(4)} (\langle 1\,2\,3\,4 \rangle \chi_5^A + \text{cyclic} ) }{\langle 1 \, 2 \, 3\,4 \rangle \langle 2 \, 3 \, 4 \,5 \rangle \langle 3 \, 4 \, 5 \, 1 \rangle \langle 4 \,5 \,1 \, 2 \rangle \langle 5 \, 1 \, 2 \, 3 \rangle}
\end{align}
whose denominator contains the five factors
\begin{align}
\{ p_1, \ldots, p_5 \} = \{ \langle 1 \, 2 \, 3\,4 \rangle , \langle 2 \, 3 \, 4 \,5 \rangle , \langle 3 \, 4 \, 5 \, 1 \rangle ,\langle 4 \,5 \,1 \, 2 \rangle ,\langle 5 \, 1 \, 2 \, 3 \rangle \}\,,
\end{align}
each of which is simply a Pl\"ucker coordinate. Evaluating these in the gauge~\eqref{eqn:zmatrix2} gives
\begin{align}
\{ p_1, \ldots, p_5 \} = \{ 1, -y^1{}_5, -y^2{}_5, -y^3{}_5, -y^4{}_5 \}
\end{align}
and evaluating the bracket~\eqref{eqn:testwedo} in this basis using~\eqref{eqn:sklybrack1} gives
\begin{align}
\Omega^{(1)}_{ij} = \{ \log p_i, \log p_j \} = \left(
\begin{array}{ccccc}
 0 & 0 & 0 & 0 & 0 \\
 0 & 0 & \frac{1}{2} & \frac{1}{2} & \frac{1}{2} \\
 0 & -\frac{1}{2} & 0 & \frac{1}{2} & \frac{1}{2} \\
 0 & -\frac{1}{2} & -\frac{1}{2} & 0 & \frac{1}{2} \\
 0 & -\frac{1}{2} & -\frac{1}{2} & -\frac{1}{2} & 0 \\
\end{array}
\right).
\end{align}
Since each entry is half-integer, the five-bracket~\eqref{eqn:nmhvinvariant} passes the bracket test.

We wrote out the steps in detail in order to illustrate the general procedure, although in this trivial case the conclusion was foregone: for $n=5$ each Pl\"ucker coordinate in~\eqref{eqn:nmhvinvariant} is frozen, so each is automatically cluster adjacent to each of the others. It is however interesting to note that if we uplift~\eqref{eqn:nmhvinvariant} by introducing trivial dependence on additional particles, this simple argument no longer applies. For example, $[1,3,5,7,9]$ still passes the bracket test even though it does not involve any frozen coordinates. The fact that the five-bracket $[i, j, k, l, m]$ passes the bracket test for any choice of indices can be understood in terms of the \emph{weak separation} criterion~\cite{LZ,OPS} for determining when two Pl\"ucker coordinates are cluster adjacent. The connection between the weak separation criterion and all Yangian invariants with $n=5k$ will be explored in~\cite{LMSV}.

\subsection{\texorpdfstring{N$^2$MHV}{NNMHV}}

The 13 rational Yangian invariants with $k=2$ are listed in Table~1 of~\cite{ArkaniHamed:2012nw} (we disregard the ninth entry in the table, which is algebraic but not rational\footnote{As mentioned in~\cite{Drummond:2018dfd}, it would be very interesting if some suitably generalized version of cluster adjacency could be found which applies to algebraic functions of momentum twistors.}). They are given by
\begin{align}
\nonumber
Y^{(2)}_1 &= [1,2,(23)\cap (456), (234) \cap (56) , 6] [2,3,4,5,6] \\
\nonumber
Y^{(2)}_2 &= [1,2,(34) \cap (567), (345) \cap (67), 7] [3,4,5,6,7] \\
\nonumber
Y^{(2)}_3 &= [1,2,3,(345) \cap (67), 7] [3,4,5,6,7] \\
\nonumber
Y^{(2)}_4 &= [1,2,3,(456) \cap (78), 8] [4,5,6,7,8] \\
\nonumber
Y^{(2)}_5 &= [1,2,3,4,8] [4,5,6,7,8] \\
\nonumber
Y^{(2)}_6 &= [1,2,3,(45) \cap (678), 8] [4,5,6,7,8] \\
Y^{(2)}_7 &= [1,2,3,(45) \cap (678), (456) \cap (78)] [4,5,6,7,8]
\label{eqn:n2yangianinv}
\\
\nonumber
Y^{(2)}_8 &= [1,2,3,4,(456) \cap (78)] [4,5,6,7,8] \\
\nonumber
Y^{(2)}_9 &= [1,2,3,4,9] [5,6,7,8,9] \\
\nonumber
Y^{(2)}_{10}&=[1,2,3,4,(567) \cap (89)] [5,6,7,8,9] \\
\nonumber
Y^{(2)}_{11}&=[1,2,3,4,(56) \cap (789)] [5,6,7,8,9] \\
\nonumber
Y^{(2)}_{12}&= \varphi \times
[1,2,3,(45) \cap (789) , (46) \cap (789)] [(45) \cap (123), (46) \cap (123), 7,8,9] \\
Y^{(2)}_{13}&=[1,2,3,4,5][6,7,8,9,10]
\nonumber
\end{align}
where
\begin{align}
\label{eq:defone}
(i j) \cap (k l m ) = Z_i \langle j\, k\, l\, m \rangle - Z_j \langle i\, k \, l\, m \rangle
\end{align}
denotes the point of intersection between the line $(ij)$ and the plane $(klm)$ in momentum twistor space. The Yangian invariant $Y^{(2)}_{12}$ has the prefactor
\begin{align}
\varphi &= \frac{\langle 4 \, 5 \, (123) \cap (789) \rangle \langle 4 \, 6 \, (123) \cap (789) \rangle}{\langle 1 \, 2 \, 3 \, 4 \rangle \langle 4 \, 7 \, 8 \, 9 \rangle \langle 5 \, 6 \, (123) \cap (789) \rangle} \, ,
\end{align}
where
\begin{align}
\label{eq:deftwo}
(i j k) \cap (lmn) = (ij) \langle k\,l\,m\,n \rangle + (jk) \langle i\, l\, m\, n \rangle + (ki) \langle j\,l\,m\,n \rangle
\end{align}
denotes the line of intersection between the planes $(ijk)$ and $(lmn)$.

Following the same procedure outlined in the previous subsection, for each Yangian invariant $Y^{(2)}_a$ listed in~\eqref{eqn:n2yangianinv} we enumerate all polynomial factors its denominator contains, and then compute the associated bracket matrix $\Omega^{(2)}_a$. Explicit results for these matrices are given in appendix~\ref{app:Appdx1}. We find that each matrix is half-integer valued, and therefore conclude that all rational $k=2$ Yangian invariants satisfy the bracket test.

\subsection{\texorpdfstring{N$^3$MHV and Higher}{NNNMHV and Higher}}

For $k>2$ it is too cumbersome, and not particularly enlightening, to write explicit formulas for each of the 977 rational Yangian invariants. We can use~\cite{Bourjaily:2012gy} to compute a symbolic formula for each Yangian invariant $Y$ in terms of the parameterization~\eqref{eqn:zmatrix2}. Then we read off the list of all polynomials in the $y^I{}_a$'s that appear in the denominator of $Y$ and compute the bracket matrix~\eqref{eqn:testwedo}. We have carried out this test for all 238 rational N$^{3}$MHV invariants with $n \le 10$ (and many invariants with $n > 10)$, and find that each one passes the bracket test. Although it is straightforward in principle to continue checking higher $n$ (and $k$) invariants, it becomes computationally prohibitive.

\acknowledgments

We are grateful to J.~Golden and A.~J.~McLeod for collaboration on~\cite{Golden:2019kks} which inspired this project, to J.~Bourjaily and C.~Fraser for helpful correspondence, and to M.~Gekhtman for enlightening explanations of Poisson brackets and cluster algebras. This work was supported in part by the US Department of Energy under contract {DE}-{SC}0010010 Task A (MS, AV), and by Simons Investigator Award \#376208 (JM, AS, AV).

\appendix

\section{\texorpdfstring{Explicit Matrices for $k=2$}{Explicit Matrices for k=2}}
\label{app:Appdx1}

Using the notation given in~\eqref{eqn:n2yangianinv}, we present here for each rational N${}^2$MHV Yangian invariant the bracket matrix of its polynomial factors:

{\tiny
\begin{align*}
\Omega_{1}^{(2)} &= \left(
\begin{array}{cccccccccc}
 0 & 0 & 1 & 1 & 0 & 0 & 0 & \frac{1}{2} & -\frac{1}{2} & -1 \\
 0 & 0 & 0 & 0 & -\frac{1}{2} & 0 & -\frac{1}{2} & \frac{1}{2} & -\frac{1}{2} & -1 \\
 -1 & 0 & 0 & -1 & -\frac{3}{2} & 0 & -\frac{1}{2} & -\frac{1}{2} & -\frac{1}{2} & -1 \\
 -1 & 0 & 1 & 0 & -\frac{3}{2} & 0 & -\frac{1}{2} & 0 & -1 & -1 \\
 0 & \frac{1}{2} & \frac{3}{2} & \frac{3}{2} & 0 & \frac{1}{2} & 0 & \frac{1}{2} & -\frac{1}{2} & -1 \\
 0 & 0 & 0 & 0 & -\frac{1}{2} & 0 & -\frac{1}{2} & 0 & 0 & 0 \\
 0 & \frac{1}{2} & \frac{1}{2} & \frac{1}{2} & 0 & \frac{1}{2} & 0 & 0 & 0 & 0 \\
 -\frac{1}{2} & -\frac{1}{2} & \frac{1}{2} & 0 & -\frac{1}{2} & 0 & 0 & 0 & -\frac{1}{2} & -\frac{1}{2} \\
 \frac{1}{2} & \frac{1}{2} & \frac{1}{2} & 1 & \frac{1}{2} & 0 & 0 & \frac{1}{2} & 0 & -\frac{1}{2} \\
 1 & 1 & 1 & 1 & 1 & 0 & 0 & \frac{1}{2} & \frac{1}{2} & 0 \\
\end{array}
\right)
&\Omega_{2}^{(2)}&= \left(
\begin{array}{cccccccccc}
 0 & 0 & 1 & 0 & 0 & 0 & 0 & -1 & -\frac{1}{2} & -\frac{1}{2} \\
 0 & 0 & 0 & -\frac{1}{2} & -\frac{1}{2} & 0 & -\frac{1}{2} & -\frac{1}{2} & -\frac{1}{2} & -\frac{1}{2} \\
 -1 & 0 & 0 & -\frac{3}{2} & -\frac{3}{2} & 0 & -\frac{1}{2} & -\frac{3}{2} & -\frac{1}{2} & -\frac{1}{2} \\
 0 & \frac{1}{2} & \frac{3}{2} & 0 & -\frac{1}{2} & \frac{1}{2} & 0 & -1 & -\frac{1}{2} & -\frac{1}{2} \\
 0 & \frac{1}{2} & \frac{3}{2} & \frac{1}{2} & 0 & \frac{1}{2} & 0 & -1 & -\frac{1}{2} & -\frac{1}{2} \\
 0 & 0 & 0 & -\frac{1}{2} & -\frac{1}{2} & 0 & -\frac{1}{2} & -\frac{1}{2} & 0 & 0 \\
 0 & \frac{1}{2} & \frac{1}{2} & 0 & 0 & \frac{1}{2} & 0 & -\frac{1}{2} & 0 & 0 \\
 1 & \frac{1}{2} & \frac{3}{2} & 1 & 1 & \frac{1}{2} & \frac{1}{2} & 0 & 0 & 0 \\
 \frac{1}{2} & \frac{1}{2} & \frac{1}{2} & \frac{1}{2} & \frac{1}{2} & 0 & 0 & 0 & 0 & -\frac{1}{2} \\
 \frac{1}{2} & \frac{1}{2} & \frac{1}{2} & \frac{1}{2} & \frac{1}{2} & 0 & 0 & 0 & \frac{1}{2} & 0 \\
\end{array}
\right)
\\
\Omega_{3}^{(2)} &= \left(
\begin{array}{cccccccccc}
 0 & 0 & \frac{1}{2} & 0 & 0 & 0 & 0 & -1 & 0 & -\frac{1}{2} \\
 0 & 0 & -\frac{1}{2} & -\frac{1}{2} & -\frac{1}{2} & 0 & -\frac{1}{2} & -\frac{1}{2} & 0 & -\frac{1}{2} \\
 -\frac{1}{2} & \frac{1}{2} & 0 & -1 & -1 & 0 & -\frac{1}{2} & -\frac{3}{2} & -\frac{1}{2} & -\frac{1}{2} \\
 0 & \frac{1}{2} & 1 & 0 & -\frac{1}{2} & \frac{1}{2} & 0 & -1 & 0 & -\frac{1}{2} \\
 0 & \frac{1}{2} & 1 & \frac{1}{2} & 0 & \frac{1}{2} & 0 & -1 & 0 & -\frac{1}{2} \\
 0 & 0 & 0 & -\frac{1}{2} & -\frac{1}{2} & 0 & -\frac{1}{2} & -\frac{1}{2} & 0 & 0 \\
 0 & \frac{1}{2} & \frac{1}{2} & 0 & 0 & \frac{1}{2} & 0 & -\frac{1}{2} & 0 & 0 \\
 1 & \frac{1}{2} & \frac{3}{2} & 1 & 1 & \frac{1}{2} & \frac{1}{2} & 0 & 0 & 0 \\
 0 & 0 & \frac{1}{2} & 0 & 0 & 0 & 0 & 0 & 0 & -\frac{1}{2} \\
 \frac{1}{2} & \frac{1}{2} & \frac{1}{2} & \frac{1}{2} & \frac{1}{2} & 0 & 0 & 0 & \frac{1}{2} & 0 \\
\end{array}
\right)
&\Omega_{4}^{(2)} &= \left(
\begin{array}{cccccccccc}
 0 & 0 & 0 & 0 & 0 & 0 & 0 & -1 & -1 & 0 \\
 0 & 0 & -\frac{1}{2} & -\frac{1}{2} & -\frac{1}{2} & 0 & -\frac{1}{2} & -\frac{1}{2} & -\frac{1}{2} & 0 \\
 0 & \frac{1}{2} & 0 & -\frac{1}{2} & -\frac{1}{2} & \frac{1}{2} & 0 & -1 & -1 & 0 \\
 0 & \frac{1}{2} & \frac{1}{2} & 0 & -\frac{1}{2} & \frac{1}{2} & 0 & -1 & -1 & 0 \\
 0 & \frac{1}{2} & \frac{1}{2} & \frac{1}{2} & 0 & \frac{1}{2} & 0 & -1 & -1 & 0 \\
 0 & 0 & -\frac{1}{2} & -\frac{1}{2} & -\frac{1}{2} & 0 & -\frac{1}{2} & -\frac{1}{2} & -\frac{1}{2} & 0 \\
 0 & \frac{1}{2} & 0 & 0 & 0 & \frac{1}{2} & 0 & -\frac{1}{2} & -\frac{1}{2} & 0 \\
 1 & \frac{1}{2} & 1 & 1 & 1 & \frac{1}{2} & \frac{1}{2} & 0 & -\frac{1}{2} & 0 \\
 1 & \frac{1}{2} & 1 & 1 & 1 & \frac{1}{2} & \frac{1}{2} & \frac{1}{2} & 0 & 0 \\
 0 & 0 & 0 & 0 & 0 & 0 & 0 & 0 & 0 & 0 \\
\end{array}
\right)
\\
\Omega_{5}^{(2)} &= \left(
\begin{array}{cccccccccc}
 0 & 0 & 0 & 0 & 0 & 0 & 0 & 0 & 0 & 0 \\
 0 & 0 & -\frac{1}{2} & -\frac{1}{2} & -\frac{1}{2} & 0 & -\frac{1}{2} & -\frac{1}{2} & -\frac{1}{2} & 0 \\
 0 & \frac{1}{2} & 0 & -\frac{1}{2} & -\frac{1}{2} & \frac{1}{2} & 0 & 0 & 0 & 0 \\
 0 & \frac{1}{2} & \frac{1}{2} & 0 & -\frac{1}{2} & \frac{1}{2} & 0 & 0 & 0 & 0 \\
 0 & \frac{1}{2} & \frac{1}{2} & \frac{1}{2} & 0 & \frac{1}{2} & 0 & 0 & 0 & 0 \\
 0 & 0 & -\frac{1}{2} & -\frac{1}{2} & -\frac{1}{2} & 0 & -\frac{1}{2} & -\frac{1}{2} & -\frac{1}{2} & 0 \\
 0 & \frac{1}{2} & 0 & 0 & 0 & \frac{1}{2} & 0 & -\frac{1}{2} & -\frac{1}{2} & 0 \\
 0 & \frac{1}{2} & 0 & 0 & 0 & \frac{1}{2} & \frac{1}{2} & 0 & -\frac{1}{2} & 0 \\
 0 & \frac{1}{2} & 0 & 0 & 0 & \frac{1}{2} & \frac{1}{2} & \frac{1}{2} & 0 & 0 \\
 0 & 0 & 0 & 0 & 0 & 0 & 0 & 0 & 0 & 0 \\
\end{array}
\right)
&\Omega_{6}^{(2)} &= \left(
\begin{array}{cccccccccc}
 0 & 0 & 0 & 0 & 0 & 0 & 0 & 0 & -1 & 0 \\
 0 & 0 & -\frac{1}{2} & -\frac{1}{2} & -\frac{1}{2} & 0 & -\frac{1}{2} & -\frac{1}{2} & -\frac{1}{2} & 0 \\
 0 & \frac{1}{2} & 0 & -\frac{1}{2} & -\frac{1}{2} & \frac{1}{2} & 0 & 0 & -1 & 0 \\
 0 & \frac{1}{2} & \frac{1}{2} & 0 & -\frac{1}{2} & \frac{1}{2} & 0 & 0 & -1 & 0 \\
 0 & \frac{1}{2} & \frac{1}{2} & \frac{1}{2} & 0 & \frac{1}{2} & 0 & 0 & -1 & 0 \\
 0 & 0 & -\frac{1}{2} & -\frac{1}{2} & -\frac{1}{2} & 0 & -\frac{1}{2} & -\frac{1}{2} & -\frac{1}{2} & 0 \\
 0 & \frac{1}{2} & 0 & 0 & 0 & \frac{1}{2} & 0 & -\frac{1}{2} & -\frac{1}{2} & 0 \\
 0 & \frac{1}{2} & 0 & 0 & 0 & \frac{1}{2} & \frac{1}{2} & 0 & -\frac{1}{2} & 0 \\
 1 & \frac{1}{2} & 1 & 1 & 1 & \frac{1}{2} & \frac{1}{2} & \frac{1}{2} & 0 & 0 \\
 0 & 0 & 0 & 0 & 0 & 0 & 0 & 0 & 0 & 0 \\
\end{array}
\right)
\\
\Omega_{7} ^{(2)} &= \left(
\begin{array}{cccccccccc}
 0 & 0 & 0 & 0 & 0 & 0 & 0 & 0 & -1 & 0 \\
 0 & 0 & -1 & -1 & -1 & 0 & 0 & -1 & -1 & 0 \\
 0 & 1 & 0 & -\frac{1}{2} & -\frac{1}{2} & \frac{1}{2} & \frac{1}{2} & -\frac{1}{2} & -\frac{3}{2} & 0 \\
 0 & 1 & \frac{1}{2} & 0 & -\frac{1}{2} & \frac{1}{2} & \frac{1}{2} & -\frac{1}{2} & -\frac{3}{2} & 0 \\
 0 & 1 & \frac{1}{2} & \frac{1}{2} & 0 & \frac{1}{2} & \frac{1}{2} & -\frac{1}{2} & -\frac{3}{2} & 0 \\
 0 & 0 & -\frac{1}{2} & -\frac{1}{2} & -\frac{1}{2} & 0 & -\frac{1}{2} & -\frac{1}{2} & -\frac{1}{2} & 0 \\
 0 & 0 & -\frac{1}{2} & -\frac{1}{2} & -\frac{1}{2} & \frac{1}{2} & 0 & -\frac{1}{2} & -\frac{1}{2} & 0 \\
 0 & 1 & \frac{1}{2} & \frac{1}{2} & \frac{1}{2} & \frac{1}{2} & \frac{1}{2} & 0 & -\frac{1}{2} & 0 \\
 1 & 1 & \frac{3}{2} & \frac{3}{2} & \frac{3}{2} & \frac{1}{2} & \frac{1}{2} & \frac{1}{2} & 0 & 0 \\
 0 & 0 & 0 & 0 & 0 & 0 & 0 & 0 & 0 & 0 \\
\end{array}
\right)
&\Omega_{8}^{(2)} &= \left(
\begin{array}{cccccccccc}
 0 & 0 & 0 & 0 & 0 & 0 & 0 & 0 & 0 & 0 \\
 0 & 0 & -1 & -1 & -1 & 0 & 0 & -1 & -1 & 0 \\
 0 & 1 & 0 & -\frac{1}{2} & -\frac{1}{2} & \frac{1}{2} & \frac{1}{2} & -\frac{1}{2} & -\frac{1}{2} & 0 \\
 0 & 1 & \frac{1}{2} & 0 & -\frac{1}{2} & \frac{1}{2} & \frac{1}{2} & -\frac{1}{2} & -\frac{1}{2} & 0 \\
 0 & 1 & \frac{1}{2} & \frac{1}{2} & 0 & \frac{1}{2} & \frac{1}{2} & -\frac{1}{2} & -\frac{1}{2} & 0 \\
 0 & 0 & -\frac{1}{2} & -\frac{1}{2} & -\frac{1}{2} & 0 & -\frac{1}{2} & -\frac{1}{2} & -\frac{1}{2} & 0 \\
 0 & 0 & -\frac{1}{2} & -\frac{1}{2} & -\frac{1}{2} & \frac{1}{2} & 0 & -\frac{1}{2} & -\frac{1}{2} & 0 \\
 0 & 1 & \frac{1}{2} & \frac{1}{2} & \frac{1}{2} & \frac{1}{2} & \frac{1}{2} & 0 & -\frac{1}{2} & 0 \\
 0 & 1 & \frac{1}{2} & \frac{1}{2} & \frac{1}{2} & \frac{1}{2} & \frac{1}{2} & \frac{1}{2} & 0 & 0 \\
 0 & 0 & 0 & 0 & 0 & 0 & 0 & 0 & 0 & 0 \\
\end{array}
\right)
\\
\Omega_{9}^{(2)} &= \left(
\begin{array}{cccccccccc}
 0 & 0 & 0 & 0 & 0 & 0 & 0 & 0 & 0 & 0 \\
 0 & 0 & -\frac{1}{2} & -\frac{1}{2} & -\frac{1}{2} & \frac{1}{2} & 0 & 0 & 0 & 0 \\
 0 & \frac{1}{2} & 0 & -\frac{1}{2} & -\frac{1}{2} & \frac{1}{2} & 0 & 0 & 0 & 0 \\
 0 & \frac{1}{2} & \frac{1}{2} & 0 & -\frac{1}{2} & \frac{1}{2} & 0 & 0 & 0 & 0 \\
 0 & \frac{1}{2} & \frac{1}{2} & \frac{1}{2} & 0 & \frac{1}{2} & 0 & 0 & 0 & 0 \\
 0 & -\frac{1}{2} & -\frac{1}{2} & -\frac{1}{2} & -\frac{1}{2} & 0 & -\frac{1}{2} & -\frac{1}{2} & -\frac{1}{2} & -\frac{1}{2} \\
 0 & 0 & 0 & 0 & 0 & \frac{1}{2} & 0 & -\frac{1}{2} & -\frac{1}{2} & -\frac{1}{2} \\
 0 & 0 & 0 & 0 & 0 & \frac{1}{2} & \frac{1}{2} & 0 & -\frac{1}{2} & -\frac{1}{2} \\
 0 & 0 & 0 & 0 & 0 & \frac{1}{2} & \frac{1}{2} & \frac{1}{2} & 0 & -\frac{1}{2} \\
 0 & 0 & 0 & 0 & 0 & \frac{1}{2} & \frac{1}{2} & \frac{1}{2} & \frac{1}{2} & 0 \\
\end{array}
\right)
&\Omega_{10}^{(2)} &= \left(
\begin{array}{cccccccccc}
 0 & 0 & 0 & 0 & 0 & 0 & 0 & 0 & 0 & 0 \\
 0 & 0 & -\frac{1}{2} & -\frac{1}{2} & -\frac{1}{2} & \frac{1}{2} & \frac{1}{2} & -\frac{1}{2} & -\frac{1}{2} & -\frac{1}{2} \\
 0 & \frac{1}{2} & 0 & -\frac{1}{2} & -\frac{1}{2} & \frac{1}{2} & \frac{1}{2} & -\frac{1}{2} & -\frac{1}{2} & -\frac{1}{2} \\
 0 & \frac{1}{2} & \frac{1}{2} & 0 & -\frac{1}{2} & \frac{1}{2} & \frac{1}{2} & -\frac{1}{2} & -\frac{1}{2} & -\frac{1}{2} \\
 0 & \frac{1}{2} & \frac{1}{2} & \frac{1}{2} & 0 & \frac{1}{2} & \frac{1}{2} & -\frac{1}{2} & -\frac{1}{2} & -\frac{1}{2} \\
 0 & -\frac{1}{2} & -\frac{1}{2} & -\frac{1}{2} & -\frac{1}{2} & 0 & -\frac{1}{2} & -\frac{1}{2} & -\frac{1}{2} & -\frac{1}{2} \\
 0 & -\frac{1}{2} & -\frac{1}{2} & -\frac{1}{2} & -\frac{1}{2} & \frac{1}{2} & 0 & -\frac{1}{2} & -\frac{1}{2} & -\frac{1}{2} \\
 0 & \frac{1}{2} & \frac{1}{2} & \frac{1}{2} & \frac{1}{2} & \frac{1}{2} & \frac{1}{2} & 0 & -\frac{1}{2} & -\frac{1}{2} \\
 0 & \frac{1}{2} & \frac{1}{2} & \frac{1}{2} & \frac{1}{2} & \frac{1}{2} & \frac{1}{2} & \frac{1}{2} & 0 & -\frac{1}{2} \\
 0 & \frac{1}{2} & \frac{1}{2} & \frac{1}{2} & \frac{1}{2} & \frac{1}{2} & \frac{1}{2} & \frac{1}{2} & \frac{1}{2} & 0 \\
\end{array}
\right)
\\
\Omega_{11}^{(2)}&= \left(
\begin{array}{cccccccccc}
 0 & 0 & 0 & 0 & 0 & 0 & 0 & 0 & 0 & 0 \\
 0 & 0 & -\frac{1}{2} & -\frac{1}{2} & -\frac{1}{2} & \frac{1}{2} & \frac{1}{2} & \frac{1}{2} & -\frac{1}{2} & -\frac{1}{2} \\
 0 & \frac{1}{2} & 0 & -\frac{1}{2} & -\frac{1}{2} & \frac{1}{2} & \frac{1}{2} & \frac{1}{2} & -\frac{1}{2} & -\frac{1}{2} \\
 0 & \frac{1}{2} & \frac{1}{2} & 0 & -\frac{1}{2} & \frac{1}{2} & \frac{1}{2} & \frac{1}{2} & -\frac{1}{2} & -\frac{1}{2} \\
 0 & \frac{1}{2} & \frac{1}{2} & \frac{1}{2} & 0 & \frac{1}{2} & \frac{1}{2} & \frac{1}{2} & -\frac{1}{2} & -\frac{1}{2} \\
 0 & -\frac{1}{2} & -\frac{1}{2} & -\frac{1}{2} & -\frac{1}{2} & 0 & -\frac{1}{2} & -\frac{1}{2} & -\frac{1}{2} & -\frac{1}{2} \\
 0 & -\frac{1}{2} & -\frac{1}{2} & -\frac{1}{2} & -\frac{1}{2} & \frac{1}{2} & 0 & -\frac{1}{2} & -\frac{1}{2} & -\frac{1}{2} \\
 0 & -\frac{1}{2} & -\frac{1}{2} & -\frac{1}{2} & -\frac{1}{2} & \frac{1}{2} & \frac{1}{2} & 0 & -\frac{1}{2} & -\frac{1}{2} \\
 0 & \frac{1}{2} & \frac{1}{2} & \frac{1}{2} & \frac{1}{2} & \frac{1}{2} & \frac{1}{2} & \frac{1}{2} & 0 & -\frac{1}{2} \\
 0 & \frac{1}{2} & \frac{1}{2} & \frac{1}{2} & \frac{1}{2} & \frac{1}{2} & \frac{1}{2} & \frac{1}{2} & \frac{1}{2} & 0 \\
\end{array}
\right)
&\Omega_{12}^{(2)} &= \left(
\begin{array}{ccccccccccccc}
 0 & 0 & 0 & 0 & 0 & 0 & 0 & 0 & 0 & 0 & 0 & 0 & 0 \\
 0 & 0 & 1 & 1 & 1 & \frac{3}{2} & \frac{3}{2} & \frac{3}{2} & \frac{3}{2} & \frac{3}{2} & \frac{3}{2} & 1 & 1 \\
 0 & -1 & 0 & -\frac{1}{2} & -\frac{1}{2} & -\frac{3}{2} & -\frac{3}{2} & -\frac{3}{2} & -\frac{1}{2} & -\frac{1}{2} & -\frac{1}{2} & -\frac{1}{2} & -\frac{1}{2} \\
 0 & -1 & \frac{1}{2} & 0 & -\frac{1}{2} & -\frac{1}{2} & -\frac{1}{2} & -\frac{1}{2} & \frac{1}{2} & \frac{1}{2} & \frac{1}{2} & 0 & -\frac{1}{2} \\
 0 & -1 & \frac{1}{2} & \frac{1}{2} & 0 & -\frac{1}{2} & -\frac{1}{2} & -\frac{1}{2} & \frac{1}{2} & \frac{1}{2} & \frac{1}{2} & \frac{1}{2} & 0 \\
 0 & -\frac{3}{2} & \frac{3}{2} & \frac{1}{2} & \frac{1}{2} & 0 & -\frac{1}{2} & -\frac{1}{2} & 2 & 2 & 2 & \frac{1}{2} & \frac{1}{2} \\
 0 & -\frac{3}{2} & \frac{3}{2} & \frac{1}{2} & \frac{1}{2} & \frac{1}{2} & 0 & -\frac{1}{2} & 2 & 2 & 2 & \frac{1}{2} & \frac{1}{2} \\
 0 & -\frac{3}{2} & \frac{3}{2} & \frac{1}{2} & \frac{1}{2} & \frac{1}{2} & \frac{1}{2} & 0 & 2 & 2 & 2 & \frac{1}{2} & \frac{1}{2} \\
 0 & -\frac{3}{2} & \frac{1}{2} & -\frac{1}{2} & -\frac{1}{2} & -2 & -2 & -2 & 0 & -\frac{1}{2} & -\frac{1}{2} & -\frac{1}{2} & -\frac{1}{2} \\
 0 & -\frac{3}{2} & \frac{1}{2} & -\frac{1}{2} & -\frac{1}{2} & -2 & -2 & -2 & \frac{1}{2} & 0 & -\frac{1}{2} & -\frac{1}{2} & -\frac{1}{2} \\
 0 & -\frac{3}{2} & \frac{1}{2} & -\frac{1}{2} & -\frac{1}{2} & -2 & -2 & -2 & \frac{1}{2} & \frac{1}{2} & 0 & -\frac{1}{2} & -\frac{1}{2} \\
 0 & -1 & \frac{1}{2} & 0 & -\frac{1}{2} & -\frac{1}{2} & -\frac{1}{2} & -\frac{1}{2} & \frac{1}{2} & \frac{1}{2} & \frac{1}{2} & 0 & -\frac{1}{2} \\
 0 & -1 & \frac{1}{2} & \frac{1}{2} & 0 & -\frac{1}{2} & -\frac{1}{2} & -\frac{1}{2} & \frac{1}{2} & \frac{1}{2} & \frac{1}{2} & \frac{1}{2} & 0 \\
\end{array}
\right)
\\
\Omega_{13}^{(2)} &= \left(
\begin{array}{cccccccccc}
 0 & 0 & 0 & 0 & 0 & 0 & 0 & 0 & 0 & 0 \\
 0 & 0 & -\frac{1}{2} & -\frac{1}{2} & -\frac{1}{2} & -\frac{1}{2} & -\frac{1}{2} & -\frac{1}{2} & -\frac{1}{2} & -\frac{1}{2} \\
 0 & \frac{1}{2} & 0 & -\frac{1}{2} & -\frac{1}{2} & -\frac{1}{2} & -\frac{1}{2} & -\frac{1}{2} & -\frac{1}{2} & -\frac{1}{2} \\
 0 & \frac{1}{2} & \frac{1}{2} & 0 & -\frac{1}{2} & -\frac{1}{2} & -\frac{1}{2} & -\frac{1}{2} & -\frac{1}{2} & -\frac{1}{2} \\
 0 & \frac{1}{2} & \frac{1}{2} & \frac{1}{2} & 0 & -\frac{1}{2} & -\frac{1}{2} & -\frac{1}{2} & -\frac{1}{2} & -\frac{1}{2} \\
 0 & \frac{1}{2} & \frac{1}{2} & \frac{1}{2} & \frac{1}{2} & 0 & -\frac{1}{2} & -\frac{1}{2} & -\frac{1}{2} & -\frac{1}{2} \\
 0 & \frac{1}{2} & \frac{1}{2} & \frac{1}{2} & \frac{1}{2} & \frac{1}{2} & 0 & -\frac{1}{2} & -\frac{1}{2} & -\frac{1}{2} \\
 0 & \frac{1}{2} & \frac{1}{2} & \frac{1}{2} & \frac{1}{2} & \frac{1}{2} & \frac{1}{2} & 0 & -\frac{1}{2} & -\frac{1}{2} \\
 0 & \frac{1}{2} & \frac{1}{2} & \frac{1}{2} & \frac{1}{2} & \frac{1}{2} & \frac{1}{2} & \frac{1}{2} & 0 & -\frac{1}{2} \\
 0 & \frac{1}{2} & \frac{1}{2} & \frac{1}{2} & \frac{1}{2} & \frac{1}{2} & \frac{1}{2} & \frac{1}{2} & \frac{1}{2} & 0 \\
\end{array}
\right)
\end{align*}
}
\!\!\noindent
Each matrix $\Omega_{i}^{(2)}$ is written in the basis $B_i$ of polynomials shown below:
{\footnotesize
\begin{align*}
B_1=&\{ \an{1,2, (23) \cap(456), (234)\cap(56)}, \an{6, 1, 2, (23)\cap(456)}, \an{(234)\cap(56), 6, 1, 2},\\
& \an{(23)\cap(456), (234)\cap(56), 6, 1}, \an{2, (23)\cap(456), (234)\cap(56), 6}, \an{2,3,4,5}, \an{6,2,3,4}, \an{5,6,2,3},\\
& \an{4,5,6,2}, \an{3,4,5,6} \},\\
B_2=&\{ \an{1,2,(34)\cap(567),(345)\cap(67)}, \an{7,1,2, (34)\cap(567)}, \an{(345)\cap(67), 7, 1, 2},  \an{(34)\cap(567),\\
& (345)\cap(67), 7, 1}, \an{2, (34)\cap(567), (345)\cap(67), 7}, \an{3,4,5,6},\an{7,3,4,5}, \an{6,7,3,4}, \an{5,6,7,3},\\
&  \an{4,5,6,7} \},\\
B_3=&\{ \an{1,2,3,(345)\cap(67)}, \an{7,1,2,3}, \an{(345)\cap(67),7,1,2}, \an{3,(345)\cap(67),7,1}, \an{2,3,(345)\cap(67),7},\\
& \an{3,4,5,6}, \an{7,3,4,5}, \an{6,7,3,4}, \an{5,6,7,3}, \an{4,5,6,7} \},\\
B_4=&\{ \an{1,2,3,(456)\cap(78)}, \an{8,1,2,3}, \an{(456)\cap(78),8,1,2}, \an{3,(456)\cap(78),8,1}, \an{2,3,(456)\cap(78),8},\\
& \an{4,5,6,7}, \an{8,4,5,6}, \an{7,8,4,5}, \an{6,7,8,4}, \an{5,6,7,8}\},\\
B_5=&\{ \an{1,2,3,4},\an{8,1,2,3}, \an{4,8,1,2}, \an{3,4,8,1}, \an{2,3,4,8}, \an{4,5,6,7}, \an{8,4,5,6}, \an{7,8,4,5}, \an{6,7,8,4},\\
& \an{5,6,7,8}\},\\
B_6=&\{ \an{1,2,3,(45)\cap(678)}, \an{8,1,2,3}, \an{(45)\cap(678),8,1,2}, \an{3,(45)\cap(678),8,1}, \an{2,3,(45)\cap(678),8},\\
& \an{4,5,6,7}, \an{8,4,5,6}, \an{7,8,4,5}, \an{6,7,8,4}, \an{5,6,7,8}\},\\
B_7=&\{ \an{1,2,3,(45)\cap(678)}, \an{(456)\cap(78),1,2,3}, \an{(45)\cap(678),(456)\cap(78),1,2},\\
& \an{3,(45)\cap(678),(456)\cap(78),1}, \an{2,3,(45)\cap(678),(456)\cap(78)}, \an{4,5,6,7}, \an{8,4,5,6}, \an{7,8,4,5},\\
& \an{6,7,8,4}, \an{5,6,7,8}\}, \\
B_8=&\{ \an{1,2,3,4},\an{(456)\cap(78),1,2,3}, \an{4,(456)\cap(78),1,2}, \an{3,4,(456)\cap(78),1}, \an{2,3,4,(456)\cap(78)},\\
& \an{4,5,6,7}, \an{8,4,5,6}, \an{7,8,4,5}, \an{6,7,8,4}, \an{5,6,7,8}\},\\
B_{9}=&\{ \an{1,2,3,4},\an{9,1,2,3}, \an{4,9,1,2}, \an{3,4,9,1}, \an{2,3,4,9}, \an{5,6,7,8}, \an{9,5,6,7}, \an{8,9,5,6},\\
& \an{7,8,9,5}, \an{6,7,8,9}\},\\
B_{10}=&\{ \an{1,2,3,4},\an{(567)\cap(89),1,2,3}, \an{4,(567)\cap(89),1,2}, \an{3,4,(567)\cap(89),1}, \an{2,3,4,(567)\cap(89)},\\
& \an{5,6,7,8}, \an{9,5,6,7}, \an{8,9,5,6}, \an{7,8,9,5}, \an{6,7,8,9}\},\\
B_{11}=&\{ \an{1,2,3,4},\an{(56)\cap(789),1,2,3}, \an{4,(56)\cap(789),1,2}, \an{3,4,(56)\cap(789),1}, \an{2,3,4,(56)\cap(789)},\\
& \an{5,6,7,8}, \an{9,5,6,7}, \an{8,9,5,6}, \an{7,8,9,5}, \an{6,7,8,9}\},\\
B_{12}=&\{ \an{1,2,3,4},\an{4,7,8,9},\an{5,6,(123)\cap(789)},\an{1,2,3,(45)\cap(789)}, \an{(46)\cap(789),1,2,3},\\
&\an{(45)\cap(789),(46)\cap(789),1,2},\an{3,(45)\cap(789),(46)\cap(789),1},\an{2,3,(45)\cap(789),(46)\cap(789)},\\
& \an{(45)\cap(123),(46)\cap(123),7,8},\an{9,(45)\cap(123),(46)\cap(123),7},\an{8,9,(45)\cap(123),(46)\cap(123)},\\
& \an{7,8,9(45)\cap(123)},\an{(46)\cap(123),7,8,9}\},\\
B_{13}=&\{ \an{1,2,3,4},\an{5,1,2,3}, \an{4,5,1,2}, \an{3,4,5,1}, \an{2,3,4,5}, \an{6,7,8,9}, \an{10,6,7,8}, \an{9,10,6,7},\\
& \an{8,9,10,6}, \an{7,8,9,10}\}\,.
\end{align*}
}
\!\!\noindent
Each basis $B_i$ is simply a list of the factors in the denominator of the Yangian invariant $Y^{(2)}_i$ given in~\eqref{eqn:n2yangianinv}.  These may be evaluated explicitly in terms of the $y^i_j$ variables by taking the appropriate maximal minors of~\eqref{eqn:zmatrix2}, using also the definitions~(\ref{eq:defone}) and~(\ref{eq:deftwo}). For example, the basis $B_1$ given above evaluates on~\eqref{eqn:zmatrix2} to
\begin{align*}
B_1=&\{p^1_1,p^1_2,...,p^1_{10}\}=\{\left(y^1_5 y^3_6-y^1_6 y^3_5\right) \left(y^1_6 y^4_5-y^1_5 y^4_6\right),y^4_6\left(y^1_6y^3_5-y^1_5 y^3_6\right) ,y^1_6\left(y^3_5 y^4_6-y^3_6 y^4_5\right),\\
&y^1_6\left(y^2_6 y^3_5-y^2_5 y^3_6\right) \left(y^1_6 y^4_5-y^1_5 y^4_6\right),y^1_6\left(y^1_6y^3_5-y^1_5 y^3_6\right) \left(y^1_6 y^4_5-y^1_5 y^4_6\right),-y^1_5,y^1_6,y^1_5 y^4_6-y^1_6y^4_5,\\
&y^1_5 y^3_6-y^1_6 y^3_5,y^1_5 y^2_6-y^1_6 y^2_5\}\,.
\end{align*}
Then one can compute the matrix $\Omega^{(2)}_{1,ij}=\{\log p^1_i,\log p^1_j\}$ simply by taking derivatives using~\eqref{eqn:sklybrack1} and the chain rule~\eqref{eqn:sklychainrule}.


\begin{thebibliography}{99}

\bibitem{Golden:2013xva}
  J.~Golden, A.~B.~Goncharov, M.~Spradlin, C.~Vergu and A.~Volovich,
  ``Motivic Amplitudes and Cluster Coordinates,''
  JHEP {\bf 1401}, 091 (2014)
  [arXiv:1305.1617 [hep-th]].

\bibitem{Golden:2013lha}
  J.~Golden and M.~Spradlin,
  ``The differential of all two-loop MHV amplitudes in \texorpdfstring{$\mathcal{N}=4$}{N=4} Yang-Mills theory,''
  JHEP {\bf 1309}, 111 (2013)
  [arXiv:1306.1833 [hep-th]].

\bibitem{Golden:2014pua}
  J.~Golden and M.~Spradlin,
  ``A Cluster Bootstrap for Two-Loop MHV Amplitudes,''
  JHEP {\bf 1502}, 002 (2015)
  [arXiv:1411.3289 [hep-th]].

\bibitem{DelDuca:2016lad}
  V.~Del Duca, S.~Druc, J.~Drummond, C.~Duhr, F.~Dulat, R.~Marzucca, G.~Papathanasiou and B.~Verbeek,
  ``Multi-Regge kinematics and the moduli space of Riemann spheres with marked points,''
  JHEP {\bf 1608}, 152 (2016)
  [arXiv:1606.08807 [hep-th]].

\bibitem{Golden:2014xqa}
  J.~Golden, M.~F.~Paulos, M.~Spradlin and A.~Volovich,
  ``Cluster Polylogarithms for Scattering Amplitudes,''
  J.\ Phys.\ A {\bf 47}, no. 47, 474005 (2014)
  [arXiv:1401.6446 [hep-th]].

\bibitem{Golden:2014xqf}
  J.~Golden and M.~Spradlin,
  ``An analytic result for the two-loop seven-point MHV amplitude in \texorpdfstring{$\mathcal{N}=4$}{N=4} SYM,''
  JHEP {\bf 1408}, 154 (2014)
  [arXiv:1406.2055 [hep-th]].

\bibitem{Harrington:2015bdt}
  T.~Harrington and M.~Spradlin,
  ``Cluster Functions and Scattering Amplitudes for Six and Seven Points,''
  JHEP {\bf 1707}, 016 (2017)
  [arXiv:1512.07910 [hep-th]].

\bibitem{Golden:2018gtk}
  J.~Golden and A.~J.~Mcleod,
  ``Cluster Algebras and the Subalgebra Constructibility of the Seven-Particle Remainder Function,''
  JHEP {\bf 1901}, 017 (2019)
  [arXiv:1810.12181 [hep-th]].

\bibitem{ArkaniHamed:2012nw}
  N.~Arkani-Hamed, J.~L.~Bourjaily, F.~Cachazo, A.~B.~Goncharov, A.~Postnikov and J.~Trnka,
  ``Grassmannian Geometry of Scattering Amplitudes,''
  arXiv:1212.5605 [hep-th].

\bibitem{Drummond:2017ssj}
  J.~Drummond, J.~Foster and {\" O}.~G{\"u}rdo{\u g}an,
  ``Cluster Adjacency Properties of Scattering Amplitudes in \texorpdfstring{$\mathcal{N}=4$}{N=4} Supersymmetric Yang-Mills Theory,''
  Phys.\ Rev.\ Lett.\  {\bf 120}, no. 16, 161601 (2018)
  [arXiv:1710.10953 [hep-th]].

\bibitem{CaronHuot:2011ky}
  S.~Caron-Huot,
  ``Superconformal symmetry and two-loop amplitudes in planar \texorpdfstring{$\mathcal{N}=4$}{N=4} super Yang-Mills,''
  JHEP {\bf 1112}, 066 (2011)
  [arXiv:1105.5606 [hep-th]].

\bibitem{CaronHuot:2011kk}
  S.~Caron-Huot and S.~He,
  ``Jumpstarting the All-Loop S-Matrix of Planar \texorpdfstring{$\mathcal{N}=4$}{N=4} Super Yang-Mills,''
  JHEP {\bf 1207}, 174 (2012)
  [arXiv:1112.1060 [hep-th]].

\bibitem{Dixon:2014iba}
  L.~J.~Dixon and M.~von Hippel,
  ``Bootstrapping an NMHV amplitude through three loops,''
  JHEP {\bf 1410}, 065 (2014)
  [arXiv:1408.1505 [hep-th]].

\bibitem{Drummond:2014ffa}
  J.~M.~Drummond, G.~Papathanasiou and M.~Spradlin,
  ``A Symbol of Uniqueness: The Cluster Bootstrap for the 3-Loop MHV Heptagon,''
  JHEP {\bf 1503}, 072 (2015)
  [arXiv:1412.3763 [hep-th]].

\bibitem{Dixon:2015iva}
  L.~J.~Dixon, M.~von Hippel and A.~J.~McLeod,
  ``The four-loop six-gluon NMHV ratio function,''
  JHEP {\bf 1601}, 053 (2016)
  [arXiv:1509.08127 [hep-th]].

\bibitem{Caron-Huot:2016owq}
  S.~Caron-Huot, L.~J.~Dixon, A.~McLeod and M.~von Hippel,
  ``Bootstrapping a Five-Loop Amplitude Using Steinmann Relations,''
  Phys.\ Rev.\ Lett.\  {\bf 117}, no. 24, 241601 (2016)
  [arXiv:1609.00669 [hep-th]].

\bibitem{Dixon:2016apl}
  L.~J.~Dixon, M.~von Hippel, A.~J.~McLeod and J.~Trnka,
  ``Multi-loop positivity of the planar \texorpdfstring{$\mathcal{N}=4$}{N=4} SYM six-point amplitude,''
  JHEP {\bf 1702}, 112 (2017)
  [arXiv:1611.08325 [hep-th]].

\bibitem{Dixon:2016nkn}
  L.~J.~Dixon, J.~Drummond, T.~Harrington, A.~J.~McLeod, G.~Papathanasiou and M.~Spradlin,
  ``Heptagons from the Steinmann Cluster Bootstrap,''
  JHEP {\bf 1702}, 137 (2017)
  [arXiv:1612.08976 [hep-th]].

\bibitem{Drummond:2018dfd}
  J.~Drummond, J.~Foster and {\" O}.~G{\"u}rdo{\u g}an,
  ``Cluster adjacency beyond MHV,''
  JHEP {\bf 1903}, 086 (2019)
  [arXiv:1810.08149 [hep-th]].

\bibitem{Drummond:2018caf}
  J.~Drummond, J.~Foster, {\" O}.~G{\"u}rdo{\u g}an and G.~Papathanasiou,
  ``Cluster adjacency and the four-loop NMHV heptagon,''
  JHEP {\bf 1903}, 087 (2019)
  [arXiv:1812.04640 [hep-th]].

\bibitem{Caron-Huot:2019vjl}
  S.~Caron-Huot, L.~J.~Dixon, F.~Dulat, M.~von Hippel, A.~J.~McLeod and G.~Papathanasiou,
  ``Six-Gluon Amplitudes in Planar \texorpdfstring{$\mathcal{N}=4$}{N=4} Super-Yang-Mills Theory at Six and Seven Loops,''
  [arXiv:1903.10890 [hep-th]].

\bibitem{Golden:2019kks}
  J.~Golden, A.~J.~McLeod, M.~Spradlin and A.~Volovich,
  ``The Sklyanin Bracket and Cluster Adjacency at All Multiplicity,''
  JHEP {\bf 1903}, 195 (2019)
  [arXiv:1902.11286 [hep-th]].

\bibitem{Steinmann}
  O.~Steinmann,
  ``\"Uber den Zusammenhang zwischen den Wightmanfunktionen und der retardierten Kommutatoren,''
  Helv.\ Phys.\ Acta {\bf 33}, 257 (1960).

\bibitem{Steinmann2}
  O.~Steinmann,
  ``Wightman-Funktionen und retardierten Kommutatoren II,''
  Helv.\ Phys.\ Acta {\bf 33}, 347 (1960).

\bibitem{Cahill:1973qp}
  K.~E.~Cahill and H.~P.~Stapp,
  ``Optical Theorems And Steinmann Relations,''
  Annals Phys.\  {\bf 90}, 438 (1975).

\bibitem{Sklyanin:1982tf}
  E.~K.~Sklyanin,
  ``Some algebraic structures connected with the Yang-Baxter equation,''
  Funct.\ Anal.\ Appl.\  {\bf 16}, 263 (1982)
  [Funkt.\ Anal.\ Pril.\  {\bf 16N4}, 27 (1982)].

\bibitem{GSV}
  M.~Gekhtman, M.~Z. Shapiro and A.~D. Vainshtein,
  ``Cluster algebras and poisson geometry,''
  Moscow Math.\ J.\ {\bf 3}, 899 (2003)
  [math/0208033].

\bibitem{Lukowski:2019sxw} 
  T.~\L ukowski, M.~Parisi, M.~Spradlin and A.~Volovich,
  ``Cluster Adjacency for $m=2$ Yangian Invariants,''
  arXiv:1908.07618 [hep-th].

\bibitem{Hodges:2009hk}
  A.~Hodges,
  ``Eliminating spurious poles from gauge-theoretic amplitudes,''
  JHEP {\bf 1305}, 135 (2013)
  [arXiv:0905.1473 [hep-th]].

\bibitem{Drummond:2008vq}
  J.~M.~Drummond, J.~Henn, G.~P.~Korchemsky and E.~Sokatchev,
  ``Dual superconformal symmetry of scattering amplitudes in \texorpdfstring{$\mathcal{N}=4$}{N=4} super-Yang-Mills theory,''
  Nucl.\ Phys.\ B {\bf 828}, 317 (2010)
  [arXiv:0807.1095 [hep-th]].

\bibitem{LZ}
  B.~Leclerc and A.~Zelevinsky,
  ``Quasicommuting families of quantum Pl\"ucker coordinates,''
  Adv.\ Math.\ Sci.\ (Kirillov's seminar), AMS Translations 181, 85 (1998).

\bibitem{OPS}
  S.~Oh, A.~Postnikov and D.~E. Speyer,
  ``Weak separation and plabic graphs,''
  Proc.\ Lond.\ Math.\ Soc.\  {\bf 110}, 721 (2015)
  [arXiv:1109.4434 [math.CO]].

\bibitem{Vergu:2015svm}
  C.~Vergu,
  ``Polylogarithm identities, cluster algebras and the \texorpdfstring{$\mathcal{N}=4$}{N=4} supersymmetric theory,''
  arXiv:1512.08113 [hep-th].

\bibitem{Sohnius:1981sn}
  M.~F.~Sohnius and P.~C.~West,
  ``Conformal Invariance in \texorpdfstring{$\mathcal{N}=4$}{N=4} Supersymmetric Yang-Mills Theory,''
  Phys.\ Lett.\  {\bf 100B}, 245 (1981).

\bibitem{Drummond:2009fd}
  J.~M.~Drummond, J.~M.~Henn and J.~Plefka,
  ``Yangian symmetry of scattering amplitudes in \texorpdfstring{$\mathcal{N}=4$}{N=4} super Yang-Mills theory,''
  JHEP {\bf 0905}, 046 (2009)
  [arXiv:0902.2987 [hep-th]].

\bibitem{Mason:2009qx}
  L.~J.~Mason and D.~Skinner,
  ``Dual Superconformal Invariance, Momentum Twistors and Grassmannians,''
  JHEP {\bf 0911}, 045 (2009)
  [arXiv:0909.0250 [hep-th]].

\bibitem{ArkaniHamed:2009vw}
  N.~Arkani-Hamed, F.~Cachazo and C.~Cheung,
  ``The Grassmannian Origin Of Dual Superconformal Invariance,''
  JHEP {\bf 1003}, 036 (2010)
  [arXiv:0909.0483 [hep-th]].

\bibitem{ArkaniHamed:2009sx}
  N.~Arkani-Hamed, J.~Bourjaily, F.~Cachazo and J.~Trnka,
  ``Local Spacetime Physics from the Grassmannian,''
  JHEP {\bf 1101}, 108 (2011)
  [arXiv:0912.3249 [hep-th]].

\bibitem{ArkaniHamed:2009dg}
  N.~Arkani-Hamed, J.~Bourjaily, F.~Cachazo and J.~Trnka,
  ``Unification of Residues and Grassmannian Dualities,''
  JHEP {\bf 1101}, 049 (2011)
  [arXiv:0912.4912 [hep-th]].

\bibitem{Drummond:2010qh}
  J.~M.~Drummond and L.~Ferro,
  ``Yangians, Grassmannians and T-duality,''
  JHEP {\bf 1007}, 027 (2010)
  [arXiv:1001.3348 [hep-th]].

\bibitem{Drummond:2010uq}
  J.~M.~Drummond and L.~Ferro,
  ``The Yangian origin of the Grassmannian integral,''
  JHEP {\bf 1012}, 010 (2010)
  [arXiv:1002.4622 [hep-th]].

\bibitem{Ashok:2010ie}
  S.~K.~Ashok and E.~Dell'Aquila,
  ``On the Classification of Residues of the Grassmannian,''
  JHEP {\bf 1110}, 097 (2011)
  [arXiv:1012.5094 [hep-th]].

\bibitem{Bourjaily:2012gy}
  J.~L.~Bourjaily,
  ``Positroids, Plabic Graphs, and Scattering Amplitudes in Mathematica,''
  arXiv:1212.6974 [hep-th].

\bibitem{Parke:1986gb}
  S.~J.~Parke and T.~R.~Taylor,
  ``An Amplitude for $n$ Gluon Scattering,''
  Phys.\ Rev.\ Lett.\  {\bf 56}, 2459 (1986).

\bibitem{Nair:1988bq}
  V.~P.~Nair,
  ``A Current Algebra for Some Gauge Theory Amplitudes,''
  Phys.\ Lett.\ B {\bf 214}, 215 (1988).

\bibitem{Drummond:2008cr}
  J.~M.~Drummond and J.~M.~Henn,
  ``All tree-level amplitudes in \texorpdfstring{$\mathcal{N}=4$}{N=4} SYM,''
  JHEP {\bf 0904}, 018 (2009)
  [arXiv:0808.2475 [hep-th]].

\bibitem{LMSV}
  L.~Lippstreu, J.~Mago, M.~Spradlin and A.~Volovich,
  ``Weak Separation, Positivity and Extremal Yangian Invariants,''
  arXiv:1906.11034 [hep-th].

\end{thebibliography}
\end{document}